\documentstyle{article}

\def\etadg{\eta^\dagger}   
\def\rhodg{\rho^\dagger}
\def\chidg{\chi^\dagger}

\def\chid{\chi^\dagger \chi}
\def\etad{\eta^\dagger \eta}
\def\rhod{\rho^\dagger \rho}
\def\sds{S^\dagger S}

\def\rt2{\frac{1}{\sqrt{2}}}

\begin{document}

\hfill{UM-P-98/52}

\vspace{30mm}
\begin{center}
{\LARGE \bf The Scalar Sector in 331 Models} 

\vspace{30mm}
{\bf M.B.Tully 
\footnote{E-mail: mbt@physics.unimelb.edu.au} 
and G.C.Joshi}
\footnote{E-mail:  joshi@bradman.ph.unimelb.edu.au}

\vspace{10mm}
{\sl 
Research Centre for High Energy Physics \\
School of Physics, University of Melbourne, \\
Parkville, Victoria 3052, Australia.}

\end{center}

\vspace{40mm}

\begin{abstract}

We calculate the exact tree-level scalar mass matrices 
resulting from
symmetry breaking using the most general gauge-invariant 
scalar 
potential of the 331 model, both with and without
the condition that lepton number is conserved.
Physical masses are also obtained in some cases,
as well as couplings to standard and exotic gauge bosons.
 
\end{abstract}

\newpage

\section{Introduction}
The 331 model \cite{B0,A0,A75} 
is an extension to the Standard Model in which the
gauge group is $SU(3)_c \times SU(3)_L \times U(1)_N$.
Some of its interesting features include that the
third family of quarks is treated differently to the
first two, that the 
number of families is required to be three (or a multiple
of three) to ensure anomaly cancellation, and the
existence of new phenomenology which would be
apparent at reasonably low energies. In particular,
a doublet of charged vector bosons carrying a lepton
number $L=2$ (known as bileptons) is predicted,
as well as an additional neutral gauge boson, $Z^\prime$.

Most analysis of the 331 model has centred on the phenomenology
of the bileptons \cite{A26,blph} and $Z^\prime$ \cite{zpph,A22}.
While certain aspects of the scalar
sector of such models have also been discussed several
times \cite{B58}-\cite{Nguyen} (including 
supersymmetric extensions \cite{A70,A52,A6})
these discussions have usually
not considered a fully general potential for the
scalar fields. The purpose of this work is therefore
to calculate the exact tree-level mass matrices
for the scalar particles using a potential which contains
all possible gauge-invariant terms, both with and
without those which violate lepton-number. In some cases
the masses of the physical states will also be given,
and coupling of the Higgs particles to 
both standard and exotic gauge bosons will also be
briefly discussed.

We will be considering the original version of the
331 model, in which each lepton
triplet is composed of $(\nu_l,l^-,l^+)_L$, where $l=e,\mu,
\tau$. \cite{exeg} The Higgs content at first consisted of three
triplets
\begin{equation}
\eta= \left( \begin{array}{c}
\eta^0 \\ \eta_1^- \\ \eta_2^+ 
\end{array} \right) 
\sim (3,0),
\rho= \left( \begin{array}{c}
\rho^+ \\ \rho^0 \\ \rho^{++}
\end{array} \right)
\sim(3,1),
\chi= \left( \begin{array}{c}
\chi^- \\ \chi^{--} \\ \chi^0 
\end{array} \right)
\sim (3,-1)
\end{equation}
however it was shown \cite{A0,B64} that in order to give the
leptons realistic masses, a scalar sextet is also needed.
\begin{equation}
S= \left( \begin{array}{ccc}
\sigma_1^0 & s_2^- & s_1^+ \\
s_2^- & S_1^{--} & \sigma_2^0 \\
s_1^+ & \sigma_2^0 & S_2^{++}
\end{array} \right)
\sim (6,0)
\end{equation}
An interesting feature of the 331 model is that
that particles within a multiplet do not all carry the
same value of lepton number, a fact which arises from
having the charged lepton from each family 
in the same triplet as its anti-particle.
The lepton number carried by the scalars can
be obtained from inspection of the Yukawa couplings
to fermions. \cite{B64}
The values obtained in this way are:
\begin{eqnarray}
L(\chi^-,\chi^{--},\sigma_1^0,s_2^-,S_1^{--}) & = & +2  \\
L(\eta_2^+,\rho^{++},S_2^{++}) & = & -2 \\
L(\eta^0,\eta_1^-,\rho^+,\rho^0,\chi^0,s_1^+,\sigma_2^0) & = & 0
\end{eqnarray}
There are thus two ways in which lepton number
can be violated in the scalar sector, either by allowing terms 
in the potential which do not conserve
the values of lepton number assigned above,
or by allowing the field 
$\sigma_1^0$ to develop a VEV, in which case
lepton number is spontaneously broken.
These possibilities will be discussed in section 5.
Until that point we will be concerned only with
the minimal model in which lepton number
is conserved.

The formalism used will be based on that of Ref. \cite{SHER}. The 
three triplets and the sextet will be made up of complex
fields $\phi_x$, each field then composed of real
fields $\phi_x = \rt2 \left( \phi_{x_1}+i\phi_{x_2} \right)$.
\begin{equation}
\eta= \rt2 \left( \begin{array}{c} 
\phi_{a_1}+i\phi_{a_2} \\
\phi_{b_1}+i\phi_{b_2} \\
\phi_{c_1}+i\phi_{c_2}
\end{array} \right),
\rho= \rt2 \left( \begin{array}{c}
\phi_{d_1}+i\phi_{d_2} \\
\phi_{e_1}+i\phi_{e_2} \\
\phi_{f_1}+i\phi_{f_2}     
\end{array} \right),
\chi= \rt2 \left( \begin{array}{c}
\phi_{g_1}+i\phi_{g_2} \\
\phi_{h_1}+i\phi_{h_2} \\
\phi_{k_1}+i\phi_{k_2} 
\end{array}
\right)
\end{equation}
\begin{equation}
S= \rt2 \left( \begin{array}{ccc}
\phi_{s_1}+i\phi_{s_2} & \frac{\phi_{p_1}+i\phi_{p_2}}{\sqrt{2}} & 
\frac{\phi_{q_1}+i\phi_{q_2}}{\sqrt{2}} \\
\frac{\phi_{p_1}+i\phi_{p_2}}{\sqrt{2}} & \phi_{t_1}+i\phi_{t_2} & 
\frac{\phi_{r_1}+i\phi_{r_2}}{\sqrt{2}} \\
\frac{\phi_{q_1}+i\phi_{q_2}}{\sqrt{2}} & \frac{\phi_{r_1}+i\phi_{r_2}}{\sqrt{2}} &
\phi_{u_1}+i\phi_{u_2}
\end{array}
\right)
\end{equation}
The mass matrices will then be calculated, using
\begin{equation}
M_{ij}^2 = \frac{\partial^2 V}{\partial \phi_i \partial \phi_j}
\end{equation}
evaluated at the chosen minimum. 
So that the particular set of values of VEVs chosen is actually a minimum,
all first derivatives
$\frac{\partial V}{\partial {\phi_x}_i}$ must be zero and
the second derivatives are required to be non-negative.
Where the mass matrices can be diagonalised, the eigenvalues
and eigenstates will also be given.

\section{Form of the Scalar Potential}
Previous studies of the scalar sector of the 331 model 
by Tonasse \cite{B23}
and recently by Nguyen, Nguyen and Long \cite{Nguyen}
have been
%In Ref. \cite{B23}, Tonasse studied the scalar sector of 331 models
based on the following potential: 
\begin{eqnarray}
V_1(\eta,\rho,\chi,S) & = & \mu_1^2 \etad + \mu_2^2 \rhod + 
\mu_3^2 \chid + \lambda_1 (\etad)^2 + \lambda_2 (\rhod)^2
\\ \nonumber
 & + & 
\lambda_3 (\chid)^2 + \lambda_4 (\etad)(\rhod) + \lambda_5
(\etad)(\chid) + \lambda_6 (\rhod)(\chid) \\ \nonumber
& + &
\lambda_7 (\rhodg \eta)(\etadg \rho) + \lambda_8 (\chidg \eta)
(\etadg \chi) + \lambda_9 (\rhodg \chi)(\chidg \rho)
\\ \nonumber
 & + & 
f_1 \left( \eta \rho \chi + H.c. \right)
\\ \nonumber
& + & 
\mu_4^2 Tr(\sds) + \lambda_{10} \left[ Tr(\sds) \right] ^2 + 
\lambda_{11}Tr \left[ (\sds)^2 \right] \\ \nonumber
& + &
\left[ \lambda_{12} (\etad) + \lambda_{13} (\rhod) + \lambda_{14}
(\chid) \right] Tr (\sds)
\\ \nonumber
& + & f_2 (\rho \chi S + H.c.)
\end{eqnarray}
where the coefficients $f_1$ and $f_2$ have dimensions of mass.

Written in a different notation, the  same potential was also 
used by Foot et al. \cite{B64} 
where the discrete symmetry
\begin{eqnarray}
\rho & \rightarrow & i \rho \\
\chi & \rightarrow & i \chi \\
\eta & \rightarrow & - \eta \\
S    & \rightarrow & -S
\end{eqnarray}
was introduced.
This was for the purpose of preventing terms which violate
lepton number from appearing such as $SSS$ and $\eta S^\dagger \eta$.
However, there are a number of additional possible
gauge invariant terms which are not excluded by
this symmetry. (The full potential is given by
Liu and Ng in Ref. \cite{A57}) 
While it would be possible to choose a different symmetry
which did prevent some of the extra terms, for example
\begin{eqnarray}
\eta & \rightarrow & - \eta \\
\rho & \rightarrow & - \rho \\
\chi & \rightarrow & i\chi \\
S    & \rightarrow & -iS
\end{eqnarray}
such choices also exclude either $f_1 \eta \rho \chi$
or $f_2 \rho \chi S$ 
which are necessary to ensure that
no continuous symmetries higher than $SU(3) \times U(1)$ exist,
and thus avoid 
the generation of additional Goldstone bosons. \cite{A22}
Therefore for this section, the original choice of 
discrete symmetry will
be retained, and the potential extended to

\begin{eqnarray}
V_2(\eta,\rho,\chi,S) & = & V_1 + 
\lambda_{15}\etadg S S^\dagger \eta + \lambda_{16} \rhodg S S^\dagger \rho
 + \lambda_{17} \chidg S S^\dagger  \chi \\ \nonumber
 & + & (\lambda_{19} \rhodg S \rho \eta + \lambda_{20} \chidg S \chi \eta
+ \lambda_{21} \eta \eta SS + H.c.)
\end{eqnarray}

Note that the $SU(3)$ invariant contractions for the last three
terms are formed by:
\begin{equation}
\begin{array}{c}
\epsilon_{ijk} \rho_l^\dagger S^{li} \rho^j \eta^k \\
\epsilon_{ijk} \chi_l^\dagger S^{li} \chi^j \eta^k \\
\epsilon_{ijk} \epsilon_{lmn} \eta^i \eta^l S^{jm} S^{kn}
\end{array}
\end{equation}
\section{Calculation of the Mass Spectrum}
The triplet Higgs fields develop VEVs as follows
\begin{equation}
\langle \eta \rangle = \left( \begin{array}{c} 
v_1 \\ 0 \\ 0
\end{array} \right),
\langle \rho \rangle = \left( \begin{array}{c}
0 \\ v_2 \\0
\end{array} \right),
\langle \chi \rangle = \left( \begin{array}{c}
0 \\ 0 \\ v_3
\end{array} \right)
\end{equation}
while for the sextet, only the $\sigma_2^0$ field develops a VEV
\begin{equation}
\langle \sigma_2^0 \rangle = v_4
\end{equation}
All VEVs are taken to be real. (The possibility of 
CP-violation arising from the scalar sector has been
investigated in detail by G\'omez Dumm \cite{A32}).

Requiring this choice of VEVs to be a minimum of the potential
leads to the 
following relations.
\begin{eqnarray}
\mu_1^2 & = & -2\lambda_1v_1^2 - \lambda_4 v_2^2 - \lambda_5 v_3^2
- f_1 \frac{v_2 v_3}{v_1}
- \lambda_{12}v_4^2 + \frac{\lambda_{19}}{\sqrt{2}} \frac{v_2^2 v_4}{v_1} \\
\nonumber
 & & - \frac{\lambda_{20}}{\sqrt{2}} \frac{v_3^2 v_4}{v_1} 
-  \lambda_{21} v_4^2 \\
\mu_2^2 & = & -2 \lambda_2 v_2^2 - \lambda_4 v_1^2 - \lambda_6 v_3^2
- f_1 \frac{v_1 v_3}{v_2}
-(\lambda_{13}+\frac{\lambda_{16}}{2}) v_4^2 
\\ \nonumber 
& & + \sqrt{2} \lambda_{19} v_1 v_4 - \frac{f_2}{\sqrt{2}} \frac{v_3 v_4}{v_2} \\
\mu_3^2 & = & -2\lambda_3 v_3^2 - \lambda_5 v_1^2 - \lambda_6 v_2^2
- f_1 \frac{v_1 v_2}{v_3} 
- (\lambda_{14}+ \frac{\lambda_{17}}{2}) v_4^2  
\\ \nonumber
& & -\sqrt{2} \lambda_{20} v_1 v_4 - \frac{f_2}{\sqrt{2}} \frac{v_2 v_4}{v_3} \\
\mu_4^2 & = & -(2 \lambda_{10} + \lambda_{11}) v_4^2 
- \lambda_{12} v_1^2 - (\lambda_{13}+ \frac{\lambda_{16}}{2}) v_2^2 
- (\lambda_{14} + \frac{\lambda_{17}}{2} ) v_3^2 \\ \nonumber
 & & + \frac{\lambda_{19}}{\sqrt{2}} \frac{v_1 v_2^2}{v_4} 
- \frac{\lambda_{20}}{\sqrt{2}} \frac{v_1 v_3^2}{v_4} - \lambda_{21} v_1^2
- f_2 \frac{v_2 v_3}{2 v_4}  
\end{eqnarray}
Using these values for the $\mu_i^2$, the mass matrices can
now be calculated.
The neutral CP-even mass matrix is the most complicated. In the
basis formed by 
$(\phi_{a_1}-v_1$,$\phi_{e_1}-v_2$, $\phi_{k_1}-v_3$,$\phi_{r_1}-v_4)$  
it may be written in the following form.

\begin{equation}
\begin{array}{c}
2 \left( \begin{array}{cccc}
m_{11}v_1^2  & m_{12}v_1v_2  & m_{13}v_1 v_3 & m_{14} v_1 v_4 \\
m_{12}v_1 v_2 & m_{22} v_2^2 & m_{23}v_2 v_3 & m_{24} v_2 v_4 \\
m_{31}v_3 v_1 & m_{32} v_2 v_3 & m_{33} v_3^2 & m_{34} v_3 v_4 \\
m_{41} v_1 v_4 & m_{42} v_4 v_2 & m_{43} v_4 v_3 & m_{44} v_4^2
\end{array}
\right)
\\
- \frac{f_1}{v_1 v_2 v_3}
\left( \begin{array}{cccc}
v_2^2 v_3^2 & v_1 v_2 v_3^2 & v_1 v_2^2 v_3 & 0 \\
v_1 v_2 v_3^2 & v_1^2 v_3^2 & v_1^2 v_2 v_3 & 0 \\
v_1 v_2^2 v_3 & v_1^2 v_2 v_3 & v_1^2 v_2^2 & 0 \\
0 & 0 & 0 & 0
\end{array}
\right) \\
- \frac{f_2}{\sqrt{2}} \frac{1}{v_2 v_3 v_4}
\left( \begin{array}{cccc}
0 & 0 & 0 & 0 \\
0 & v_3^2 v_4^2 & v_2 v_3 v_4^2 & - v_2 v_3^2 v_4 \\
0 & v_2 v_3 v_4^2 & v_2^2 v_4^2 & - v_2^2 v_3 v_4 \\
0 & -v_2 v_3^2 v_4 & -v_2^2 v_3 v_4 & v_2^2 v_3^2 
\end{array}
\right)
\\   
+ \left( 
\frac{\lambda_{19}}{\sqrt{2}} \frac{v_2^2}{v_1 v4} - 
\frac{\lambda_{20}}{\sqrt{2}} \frac {v_3^2}{v_1 v_4}
-2 \lambda_{21} 
\right)
\left( \begin{array}{cccc}
v_4^2 & 0 & 0 & - v_1 v_4 \\
0 & 0 & 0 & 0 \\
0 & 0 & 0 & 0 \\
- v_1 v_4 & 0 & 0 & v_1^2
\end{array}
\right)
\end{array} \label{cpeven}
\end{equation}
where
\begin{eqnarray}
m_{11} & = & 2\lambda_1 \\
m_{12}=m_{21} & = & \left( \lambda_4 - \frac{\lambda_{19}}{\sqrt{2}}\frac{v_4}{v_1} \right)
\\
m_{13}=m_{31} & = & \left( \lambda_5 + \frac{\lambda_{20}}{\sqrt{2}} \frac{v_4}{v_1} \right)
\\
m_{14}=m_{41} & = & \left( \lambda_{12} + \lambda_{21} \right) \\
m_{22} & = & 2 \lambda_2 v_2^2 \\
m_{23}=m_{32} & = &  \lambda_6 \\
m_{24}=m_{42} & = &  \left( \lambda_{13} + \frac{\lambda_{16}}{2} - \frac{\lambda_{19}}{\sqrt{2}} 
\frac{v_1}{v_4} \right) \\
m_{33} & = &  2 \lambda_3  \\
m_{34}=m_{43} & = &   \left( \lambda_{14} + \frac{\lambda_{17}}{2} + 
\frac{\lambda_{20}}{\sqrt{2}} \frac{v_1}{v_4} \right) \\
m_{44} & = &  \left( 2\lambda_{10} + \lambda_{11} \right) 
\end{eqnarray}

This matrix will lead to the existence of four neutral
CP-even particles, denoted $H_{1,2,3,4}^0$, the masses
of which can only be obtained by making some fairly
severe approximations.

The neutral CP-odd matrix takes the simpler form:
\begin{equation}
\begin{array}{c}
- \frac{f_1}{v_1 v_2 v_3}
\left( \begin{array}{cccc}
v_2^2 v_3^2 & v_1 v_2 v_3^2 & v_1 v_2^2 v_3 & 0 \\
v_1 v_2 v_3^2 & v_1^2 v_3^2 & v_1^2 v_2 v_3 & 0 \\
v_1 v_2^2 v_3 & v_1^2 v_2 v_3 & v_1^2 v_2^2 & 0 \\
0 & 0 & 0 & 0
\end{array}
\right) \\
- \frac{f_2}{\sqrt{2}} \frac{1}{v_2 v_3 v_4}
\left( \begin{array}{cccc}
0 & 0 & 0 & 0 \\
0 & v_3^2 v_4^2 & v_2 v_3 v_4^2 & - v_2 v_3^2 v_4 \\
0 & v_2 v_3 v_4^2 & v_2^2 v_4^2 & - v_2^2 v_3 v_4 \\
0 & -v_2 v_3^2 v_4 & -v_2^2 v_3 v_4 & v_2^2 v_3^2 
\end{array}
\right)
\\   
+ \left( 
\frac{\lambda_{19}}{\sqrt{2}} \frac{v_2^2}{v_1 v4} - 
\frac{\lambda_{20}}{\sqrt{2}} \frac {v_3^2}{v_1 v_4}
-2 \lambda_{21} 
\right)
\left( \begin{array}{cccc}
v_4^2 & 0 & 0 & v_1 v_4 \\
0 & 0 & 0 & 0 \\
0 & 0 & 0 & 0 \\
v_1 v_4 & 0 & 0 & v_1^2
\end{array}
\right)
\end{array} \label{cpodd}
\end{equation}
and this leads to two massless particles $G_1^0,G_2^0$
and two physical ones, $A_1^0,A_2^0$.
There is further an additional CP-even and CP-odd state,
degenerate in mass, resulting from the $\phi_s$ field,
which carry a lepton number $L=2$, and thus do not mix
with the other neutral states.
\begin{eqnarray}
H_5^0 & = & \phi_{s_1} \\ 
A_3^0 & = & \phi_{s_2}  
\end{eqnarray}
\begin{eqnarray}
m_{H_5^0}^2 = m_{A_3^0}^2 &  =  &
-\lambda_{11}v_4^2 + \lambda_{15} v_1^2 - \frac{f_2}{\sqrt{2}} \frac{v_2 v_3}{v_4}
-\frac {\lambda_{16}}{2} v_2^2 - \frac{\lambda_{17}}{2}v_3^2 \label{ms}  
\\ \nonumber 
 & & 
+ \frac {\lambda_{19}}{\sqrt{2}} \frac {v_1 v_2^2}{v_4} - \frac {\lambda_{20}}{\sqrt{2}}
\frac{v_1 v_3^2}{v_4} - \lambda_{21} v_1^2   
\end{eqnarray}
In the singly-charged sector, there is mixing between
the $L=0$ fields 
$\phi_b$, $\phi_d^\star$ and $\phi_q^\star$ and also between
the $L=-2$ fields
$\phi_c$, $\phi_g^\star$ and $\phi_p^\star$.
The mixing between $\phi_{b}$, $\phi_{d}^\star$ and
$\phi_{q}^\star$ is given by:
\begin{equation}
l_1
\left( \begin{array}{ccc}
v_2^2 & v_1 v_2 & 0 \\
v_1 v_2 & v_1^2 & 0 \\
0 & 0 & 0 
\end{array} \right)
+ l_2 
\left( \begin{array}{ccc}
v_4^2 & 0 & v_1 v_4 \\
0 & 0 & 0 \\
v_1 v_4 & 0 & v_1^2
\end{array} \right)
+ l_3
\left( \begin{array}{ccc}
0 & 0 & 0 \\
0 & v_4^2 & - v_2 v_4 \\
0 & -v_2 v_4 & v_2^2
\end{array} \right) \label{mbdp} 
\end{equation} 
where
\begin{eqnarray}
l_1 & = & \lambda_7 -f_1 \frac{v_3}{v_1 v_2} + 
\lambda_{19} \frac{v_4}{v_1} \\
l_2 & = & \lambda_{15} -4 \lambda_{21} - 
\lambda_{20} \frac {v_3^2}{v_1 v_4} \\
l_3 & = & - \lambda_{16} - f_2 \frac{v_3}{v_2 v_4} 
+ \lambda_{19} \frac{v_1}{v_4}
\end{eqnarray}
which leads to a massless scalar $G_1^+$ and two
physical particles of mass:
\begin{eqnarray}
m_{H^+_{1,2}}^2 & = & l_1 (v_1^2 + v_2^2) + l_2 (v_1^2 + v_4^2)
+ l_3 (v_2^2+v_4^2) \\ \nonumber
& &  \pm
\frac{1}{2}
\left\{ \left[ l_1(v_1^2+v_2^2)+l_2(v_1^2+v_4^2) 
+l_3(v_2^2+v_4^2)
\right] ^2  \right. \\ \nonumber
& & 
\left. 
-4(v_1^2+v_2^2+v_4^2)(l_1l_2v_1^2 + l_1l_3v_2^2 + l_2l_3v_4^2) \right\}^{\frac{1}{2}}
\end{eqnarray}
The mixing between  $\phi_{c}$, $\phi_{g}^\star$ and $\phi_{p}^\star$
is
\begin{equation}
l_4 \left( \begin{array}{ccc}
v_3^2 & v_1 v_3 & 0 \\
v_1 v_3 & v_1^2 & 0 \\
0 & 0 & 0
\end{array} \right)
+ l_5
\left( \begin{array}{ccc}
v_4^2 & 0 & v_1 v_4 \\
0 & 0 & 0 \\
v_1 v_4 & 0 & v_1^2 
\end{array} \right)  
+ l_6
\left( \begin{array}{ccc}
0 & 0 & 0 \\
0 & v_4^2 & -v_3 v_4 \\
0 & -v_3 v_4 & v_3^2
\end{array} \right) \label{mcgq} 
\end{equation} 
where
\begin{eqnarray}
l_4 & = & \lambda_8 - f_1 \frac{v_2}{v_1 v_3} 
- \lambda_{20} \frac{v_4}{v_1} \\
l_5 & = & \lambda_{15} + \lambda_{19} \frac{v_2^2}{v_1 v_4}
- 4 \lambda_{21} \\
l_6 & = & -f_2 \frac{v_2}{v_3 v_4} + \lambda_{17} 
+ \lambda_{20} \frac{v_1}{v_4}
\end{eqnarray}
again leading to a massless state $G^+_2$ and two physical
particles 
\begin{eqnarray}
m^2_{H^+_{3,4}} & = & l_4 (v_1^2+v_3^2)+l_5(v_1^2+v_4^2)+   
l_6(v_3^2+v_4^2)  \\ \nonumber
& & \pm \frac{1}{2} \left\{ \left[ l_4(v_1^2+v_3^2) 
+l_5(v_1^2+v_4^2)+l_6(v_3^2+v_4^2)
\right]^2 \right. \\ \nonumber
& & \left. \left. -4 (v_1^2 + v_3^2 + v_4^2) (l_4l_5 v_1^2 + l_4l_6 v_3^2
+ l_5l_6 v_4^2) \right. \right\}^{\frac{1}{2}}
\end{eqnarray}
Lastly, mixing between the doubly-charged fields is given
in the $\phi_{f}$,$\phi_{h}^\star$,$\phi_{t}^\star$,$\phi_{u}$
basis, by:
\begin{equation}
\begin{array}{c}
l_7 \left( \begin{array}{cccc}
v_3^2 & v_2v_3 & 0 & 0 \\
v_2 v_3 & v_2^2 & 0 & 0 \\
0 & 0 & 0 & 0 \\
0 & 0 & 0 & 0 
\end{array} \right)
+ l_8 \left( \begin{array}{cccc}
\sqrt{2}v_4^2 & 0 & v_2 v_4 & 0 \\
0 & 0 & 0 & 0 \\
v_2 v_4 & 0 & \frac{1}{\sqrt{2}} v_2^2 & 0 \\
0 & 0 & 0 & 0 
\end{array} \right) \\
+ l_9 \left( \begin{array}{cccc}
\sqrt{2} v_4^2 & 0 & 0 & -v_2 v_4 \\
0 & 0 & 0 & 0 \\
0 & 0 & 0 & 0 \\
-v_2 v_4 & 0 & 0 & \rt2 v_2^2
\end{array} \right)
+ l_{10} \left( \begin{array}{cccc}
0 & 0 & 0 & 0 \\
0 & \sqrt{2} v_4^2 & -v_3 v_4 & 0 \\
0 & -v_3 v_4 & \rt2 v_3^2 & 0 \\
0 & 0 & 0 & 0 
\end{array} \right) \\
+ l_{11}  \left( \begin{array}{cccc}
0 & 0 & 0 & 0 \\
0 & \sqrt{2} v_4^2 & 0 & v_3 v_4 \\
0 & 0 & 0 & 0 \\
0 & v_3 v_4 & 0 & \rt2 v_3^2    
\end{array} \right)
+ l_{12} \left( \begin{array}{cccc}
0 & 0 & 0 & 0 \\
0 & 0 & 0 & 0 \\
0 & 0 & v_4^2 & v_4^2 \\
0 & 0 & v_4^2 & v_4^2 
\end{array} \right)
\end{array}
\end{equation}
where
\begin{eqnarray}
l_7 & = & \left( \lambda_9 - f_1 \frac{v_2 v_3}{v_1} +
\frac{f_1}{\sqrt{2}} \frac{v_4}{v_2 v_3} \right)\\
l_8 & = & \left( \frac{\lambda_{16}}{\sqrt{2}} +
\lambda_{19}\frac{v_1}{v_4} \right) \\
l_9 & = & \left( - \frac{\lambda_{16}}{\sqrt{2}} + 
\lambda_{19} \frac{v_1}{v_4}-f_2 \frac{v_3}{v_2 v_4}
\right) \\
l_{10} & = & \left( - \frac{\lambda_{17}}{\sqrt{2}}
- \lambda_{20} \frac{v_1}{v_4} - f_2 \frac{v_2}{v_3 v_4} 
\right) \\
l_{11} & = & \left( \frac{\lambda_{17}}{\sqrt{2}}
- \lambda_{20} \frac{v_1}{v_4} \right) \\
l_{12} & = & \left( \lambda_{11} - \lambda_{21}
\frac{v_1^2}{v_4^2} \right)
\end{eqnarray}
This matrix has zero determinant and thus there will
exist a doubly charged Goldstone boson of each sign,
$G^{++}$ and three massive particles $H_1^{++}$,$H_2^{++}$
and 
$H_3^{++}$, all having a lepton number $L=-2$.

To summarise, symmetry breaking will lead to eight Goldstone
bosons, $G_1^0,G_2^0,G_1^{\pm},G_2^{\pm}$ and $G^{\pm \pm}$,
which become incorporated into the $Z^0$,$Z^{\prime 0}$,
$W^\pm$,$Y^\pm$ and $Y^{\pm \pm}$ gauge bosons respectively.
The physical states consist of
five CP-even neutral scalars
$H_1^0,H_2^0,H_3^0,H_4^0,H_5^0$, three neutral CP-odd states
$A_1^0,A_2^0,A_3^0$, four singly-charged particles of each charge
$H_1^{\pm},H_2^{\pm},H_3^{\pm},H_4^{\pm}$ 
and three doubly-charged ones $H_1^{\pm \pm}, H_2^{\pm \pm},H_3^{\pm \pm}$.
\section{Couplings to Gauge Bosons}
After the breaking of $SU(3)_L \times U(1)_N$ symmetry the
gauge bosons are formed as follows: 
\begin{eqnarray}
\gamma & = & \sin \theta_W W_3 - \cos \theta_W \left(
\sin \varphi W_8 - \cos \varphi B \right)\\
Z & = & -\cos \theta_W W_3 - \sin \theta_W \left(
\sin \varphi W_8 - \cos \varphi B \right) \\
Z^{\prime} & = & \cos \varphi W_8 + \sin \varphi B    \\ 
W^+ & = & \rt2 (W_1-iW_2) \\
Y^- & = & \rt2 (W_4 - iW_5) \\
Y^{--} & = & \rt2 (W_6 - iW_7) 
\end{eqnarray}
where $W_a$ and $B$ represent the gauge fields of $SU(3)_L$
and $U(1)_N$ respectively, and
\begin{eqnarray}
\sin \theta_W = \frac{{g^\prime}}{\sqrt{g^2+{g^\prime}^2}}
& = & \frac{g_N}{\sqrt{g^2+4g_N^2}} \\
\sin \varphi =\sqrt{3} \tan \theta_W & = & \frac{ \sqrt{3} g_N}{\sqrt{g^2+3g_N^2}} 
\end{eqnarray}
The three coupling constants are related to each other by
\begin{equation}
\frac{1}{{g^\prime}^2}=\frac{1}{g_N^2}+ \frac{3}{g^2} 
\end{equation}
but it should be noted that this relation will clearly be
altered if the covariant derivative (eq. \ref{covder}) is written with 
an extra factor multiplying $g_N$. 
Compared to the Standard Model,
there is now an extra neutral gauge boson ($Z^\prime$) and
the two bileptons ($Y^-$,$Y^{--}$).

Couplings of the gauge bosons to the various Higgs states can
then be calculated using these definitions and the covariant
derivatives for the triplets:
\begin{equation}
D_{\mu} \varphi =\partial_{\mu}\varphi+ \frac{ig}{2}\left[ {\bf W_\mu \cdot \lambda}
\right]\varphi + ig_N N_\varphi B_\mu \varphi \label{covder}  
\end{equation} 
(where $\varphi=\eta,\rho,\chi$) and the sextet:
\begin{equation}
D_{\mu}S=\partial_{\mu}S+\frac{ig}{2} \left( \left[
{\bf W_\mu \cdot \lambda } \right] S + S \left[
{\bf W_{\mu} \cdot \lambda} \right] ^T \right)
\end{equation}
where ${\bf W_{\mu} \cdot \lambda}= W_{\mu}^a \lambda^a$, with
$Tr[\lambda^a \lambda^b]=2\delta^{ab}$. 

Considering the CP-even states, all four physical particles
are composed of mixtures of $\phi_{a_1}-v_1$,$\phi_{e_1}-v_2$,
$\phi_{k_1}-v_3$ and $\phi_{r_1}-v_4$. However, since in the 331 model
$v_3 > v_1,v_2,v_4$ it can be assumed that the lightest
state, $H_1^0$ is made up predominantly of 
$\phi_{a_1}-v_1$,$\phi_{e_1}-v_2$, and
$\phi_{k_1}-v_4$ only. 

Using this approximation, the following quartic couplings
are found.
\begin{eqnarray}
g \left( H_1^0 H_1^0 \gamma \gamma \right) & = & 0 \\
g \left( H_1^0 H_1^0 Z Z \right) & \simeq & \frac{1}{2} \left(
g^2 + {g^\prime}^2 \right) \\
g \left( H_1^0 H_1^0 W^+ W^- \right) & \simeq & \frac{g^2}{2}
\end{eqnarray}
This state therefore resembles the 
minimal Standard Model Higgs boson.

Some other values for Higgs-gauge boson couplings have also
been calculated.
\begin{eqnarray}
g \left( H_1^+ H_1^- ZZ \right)  =  g \left( H_2^+ H_2^- ZZ \right) & = &
\frac{1}{2} 
\frac{(g^2 - {g^\prime}^2 )}{g^2+{g^\prime}^2}^2 \\
g \left( H_1^+ H_1^- W^+W^- \right)  =  g \left( H_2^+ H_2^- W^+W^- \right) & = & 
\frac{g^2}{2} \\
g \left( H_3^+ H_3^- Y^+Y^- \right)  =  g \left( H_4^+ H_4^- Y^+Y^- \right)
 & = & \frac{g^2}{2}
\end{eqnarray}

\section{Lepton number violation}
Until this point we have been considering a
minimal version of the 331 model in which
lepton number is conserved. 
There are two ways of extending the model so
that this is no longer the case.
Firstly, $\sigma_1^0$ can develop a VEV (denoted $v_5$)
satisfying
\begin{eqnarray}
v_5^2 & = & v_4^2-\frac{1}{\lambda_{11}} \left[
\lambda_{15}v_1^2+\lambda_{16}v_2^2
-  \frac{\lambda_{17}}{2}  v_3^2 
+ \frac{\lambda_{19}}{\sqrt{2}} \frac{v_1 v_2^2}{v_4} \right. \\ \nonumber
& & \left. 
- \frac{\lambda_{20}}{\sqrt{2}} \frac{v_1 v_3^2}{v_4} 
- \lambda_{21} v_1^2
- f_2 \frac{v_2 v_3}{2 v_4}  
\right]
\end{eqnarray}
In this case, lepton number is spontaneously broken,
and there therefore arises an additional Goldstone
boson - the majoron. In the notation used here,
the field $\phi_{s_2}$ becomes massless if 
$\langle \phi_s \rangle = v_5$.
This situation, however, is equivalent to the
well-known triplet majoron model, which is ruled
out experimentally by Z lineshape measurements,
in that the existence of such a particle would
contribute to the invisible decay width of the Z
and this is not observed. \cite{invisible}

The other possibility is to extend the potential to
include terms which,  
although gauge-invariant,
do not conserve lepton number, that is
\begin{eqnarray}
V_3(\eta,\rho,\chi,S) & = & V_2+
f_3 \eta S^\dagger \eta + f_4 SSS + 
\lambda_{22} \chi^\dagger \eta \rho^\dagger \eta \\ \nonumber & & + 
\lambda_{23} \eta^\dagger S \chi \rho + 
\lambda_{24} \chi \rho SS
+H.c.
\end{eqnarray}
The $SU(3)$ contractions for these terms are formed by
\begin{equation}
\begin{array}{c}
\eta^i S^\dagger_{ij} \eta^j \\
\epsilon_{ijk} \epsilon_{lmn}S^{il}S^{jm}S^{kn} \\
\chi^\dagger_i \eta^i \rho^\dagger_j \eta^j \\
\epsilon_{ijk} \eta^\dagger_l S^{li} \chi^j \rho^k \\
\epsilon_{ijk} \epsilon_{lmn} \chi^i \rho^l S^{jm} S^{kn}
\end{array}
\end{equation}
It is still possible
with this potential to maintain $v_5=0$ if the
condition
\begin{equation}
f_3v_1^2-3f_4v_4^2+\lambda_{23}v_1v_2v_3-\sqrt{2}\lambda_{24}v_2v_3v_4
=0
\end{equation}
is imposed. In this scenario, neutrinos are massless
at tree-level, but
develop a majorana mass at the one loop level. \cite{A66}
With regard to the scalar sector, 
the effect of the extra terms is to lead to
mixing between particles of different lepton number.
 
Previously,
the charged scalars made up of the $L=0$ fields
$\phi_b$,$\phi_d$ and $\phi_q$ did not mix with
those from the $|L|=2$ fields $\phi_c$,$\phi_g$
and $\phi_p$. Now however, mixing between the
two arises, which in the $(\phi_b,\phi_d^\star,\phi_q^\star,\phi_c^\star,
\phi_g,\phi_p)$ basis
may be written as    
\begin{equation}
\left(
\begin{array}{cc}
M_{bdq}^2 & M_{bdq-cgp}^2 \\
M_{bdq-cgp}^2 & M_{cgp}^2
\end{array}
\right)
\end{equation}
where $M^2_{bdp}$ is given by equation \ref{mbdp}, $M^2_{cgq}$
by equation \ref{mcgq}, and the new mixing terms $M_{bdq-cgp}^2$ by
\begin{equation}
\left(   
\begin{array}{ccc}
\sqrt{2}f_3v_4+\lambda_{22}v_2v_3 &
\lambda_{22}v_1v_2 - \frac{\lambda_{23}}{\sqrt{2}} v_2v_4 &
\sqrt{2}f_3v_1 + \frac{\lambda_{23}}{\sqrt{2}}v_2v_3 \\
\lambda_{22}v_1v_3 - \frac{ \lambda_{23}}{\sqrt{2}}v_3v_4 &
\lambda_{22}v_1^2 - \lambda_{24}v_4^2 &
-\frac{\lambda_{23}}{\sqrt{2}}v_1v_3 + \lambda_{24}v_3v_4 \\
\sqrt{2}f_3v_1 + \frac{\lambda_{23}}{\sqrt{2}}v_2v_3 &
-\frac{\lambda_{23}}{\sqrt{2}}v_1v_2 + \lambda_{24}v_2 v_4 &
3\sqrt{2}f_4v_4+ \lambda_{24}v_2v_3
\end{array}
\right)
\end{equation}
Similarly, with the minimal potential $V_2$
the neutral scalars arising from the $L=0$ fields
$\phi_a$,$\phi_e$ and $\phi_k$ did not mix
with those from the $L=2$ $\phi_s$ field, but
now mixing occurs between both the CP-even states
and the CP-odd.

In the CP-even basis 
($\phi_{a_1}-v_1,\phi_{e_1}-v_2,\phi_{k_1}-v_3,\phi_{r_1}-v_4,\phi_{s_1}$)
the mass matrix takes the form
\begin{equation}
\left(
\begin{array}{cc}
M_{aekr}^2 & M_{aekr-s}^2 \\
M_{aekr-s}^2 & M_s^2
\end{array}
\right)
\end{equation}
where $M_{aekr}^2$ is given by equation \ref{cpeven}, 
$M_s^2$ by equation \ref{ms}
and the mixing term $M_{aekr-s}^2$ by
\begin{equation}
\left(
\begin{array}{c}
f_3v_1 + \frac{\lambda_{23}}{\sqrt{2}}v_2v_3 \\
\frac{\lambda_{23}}{\sqrt{2}}v_1v_3-\sqrt{2}\lambda_{24}v_3v_4 \\
\frac{\lambda_{23}}{\sqrt{2}}v_1v_2-\sqrt{2}\lambda_{24}v_2v_4 \\
-6f_4v_4 - \sqrt{2}\lambda_{24}v_2v_3
\end{array}
\right)
\end{equation}
In the CP-odd basis
($\phi_{a_2},\phi_{e_2},\phi_{k_2},\phi_{r_2},\phi_{s_2}$)
the mass matrix becomes
\begin{equation}
\left(
\begin{array}{cc}
M_{aekr}^{\prime 2} & M_{aekr-s}^{\prime 2} \\
M_{aekr-s}^{\prime 2} & M_s^{\prime 2}
\end{array}
\right)
\end{equation}
where $M_{aekr}^{\prime 2}$ is given by equation \ref{cpodd}
, $M_s^{\prime 2}$ by equation \ref{ms}
and the mixing term $M_{aekr-s}^{\prime 2}$ by
\begin{equation}
\left(
\begin{array}{c}
f_3v_1 + \frac{\lambda_{23}}{\sqrt{2}}v_2v_3 \\
-\frac{\lambda_{23}}{\sqrt{2}}v_1v_3 + \sqrt{2}\lambda_{24}v_3v_4 \\
-\frac{\lambda_{23}}{\sqrt{2}}v_1v_2 + \sqrt{2}\lambda_{24}v_2v_4 \\
6f_4v_4 + \sqrt{2}\lambda_{24}v_2v_3
\end{array}
\right)
\end{equation}

The additional terms have no effect on the doubly-charged
scalars.

\section{Conclusion}
A scalar potential for the 331 model more general than those 
previously considered in detail has been studied and the scalar
mass matrices resulting from symmetry breaking calculated.
In some cases, the exact masses for the physical states
have been obtained, and the couplings to gauge bosons
calculated.
Such a potential leads to a very wide variety of 
scalar bosons.

Extensive studies have been made regarding the possible
discovery of various types of Higgs bosons at future
collider experiments. \cite{snow} The most studied have been the
CP-even $H^0$, which occurs in the Standard Model (SM),
and those states which occur in the Minimal Supersymmetric 
Standard Model (MSSM), $H^0_1$,$H^0_2$,$A^0$ and $H^{\pm}$.

A mass limit of 88 GeV has been obtained \cite{ep98144,ep98173} 
for the
SM $H^0$ by searching for the process 
$e^+e^- \rightarrow ZH^0$ at LEP, while limits of 71 GeV 
have
been given for both the lightest CP-even state $H_1^0$
and CP-odd state $A^0$ of the MSSM. \cite{ep9872} 
The current limit for $H^\pm$ is 60 GeV. \cite{ep98173,ep98149} 
Doubly-charged states have also recieved some attention.
\cite{acton}

 The phenomenology of both scalar
and vector bileptons
has been reviewed by Cuypers and Davidson \cite{A26}
who  noted that the mass of neutral as well as
 singly and
doubly-charged scalar bileptons  
are all constrained by their non-observation in
Z decay to be greater than 45 GeV. Future $e^-e^-$ and
$e^+e^-$ collider experiments give the best hope of
detecting these particles, providing their mass is
less than 
half the centre-of-mass
energy of the collider.

Clearly the phenomenology of the scalar sector of the
331 model is very rich and worthy of further study.

\end{document}